\documentclass[twocolumn,showpacs,preprintnumbers,amsmath,amssymb]{revtex4}

\usepackage{graphicx}
\usepackage{dcolumn}
\usepackage{bm}
\usepackage{times}

\begin{document}

\title{Spin Pseudo Gap in La$_{2-x}$Sr$_x$CuO$_4$ Studied by Neutron
Scattering}

\author{C. H. Lee}
\affiliation{National Institute of Advanced Industrial Science
and Technology, 1-1-1 Umezono, Tsukuba, Ibaraki 305-8568, Japan}
\author{K. Yamada}
\affiliation{Institute for Chemical Research, Kyoto University, Uji
611-0011, Japan}
\author{H. Hiraka}
\author{C. R. Venkateswara Rao}
\author{Y. Endoh}
\affiliation{Institute for Materials Research, Tohoku University, Sendai 980-8577,
Japan}
\date{\today}

\begin{abstract}
Spin excitations of $\rm La_{2-x}Sr_xCuO_4$ have been studied using
inelastic neutron scattering
techniques in the energy range of 2 meV $\leq \omega \leq$ 12 meV and the
temperature range of 8 K $\leq$ $T$ $\leq$ 150 K.
We observed a signature of a spin pseudo gap in the excitation
spectrum above $T_c$ for the
slightly overdoped sample with x = 0.18.  On heating, the spin pseudo gap
gradually collapses between $T$ = 80 K and 150 K.  
For the x = 0.15 and 0.20, although the visibility
of gap-like structure at $T$ $\sim$ $T_c$ is lower compared to the x = 0.18 
sample, the
broad bump of $\chi^{\prime \prime} (\omega)$ appears at $\omega$ 
$\sim$ 5 meV,
close to the spin-gap energy at base temperature, suggests the existence of 
the spin
pseudo gap in the normal state.
\end{abstract}


\maketitle

\section{Introduction}
\label{sec:level1}
One of the most remarkable phenomena observed in high-$T_c$ cuprates is the
opening of a pseudo gap
above the superconducting (SC) transition temperature ($T_c$) in
excitations of charge as well as
spin.  To elucidate the basis of this effect, many experimental studies have been
carried out using various 
techniques including photoemission spectroscopy [1-5], NMR [6,7], neutron
scattering [8] and others.  Nevertheless,
the microscopic origin of the pseudo gap remains controversial.  Many
theoretical models have 
been proposed based on preformed Cooper pairs or SC pairing fluctuations
[9], the RVB state [10,11],
a precursor to a spin-density-wave-state [12] and the formation of
dynamical charge stripes.
To reduce the considerable theoretical and experimental confusion regarding
the basis of the spin
pseudo gap, further experimental studies are required.

In principle, pseudo gap in magnetic excitations, so-called spin pseudo
gap, is observable using NMR
as well as neutron scattering.  Especially, neutron scattering spectroscopy has the
unique benefit of being able
to detect directly the energy gap as well as the momentum dependence of
spin fluctuation.  In fact,
neutron scattering measurements on the $\rm YBa_{2}Cu_{3}O_{6+y}$ (YBCO)
system observed for the first time a gap-like
structure in the energy spectrum of dynamical spin susceptibility
$\chi^{\prime\prime}({\bf q},\omega)$ near the ($\pi ,\pi$) position in
the normal state [8].  For underdoped $\rm La_{2-x}Sr_xCuO_4$ (LSCO),
however, no evidence for the existence of a
pseudo gap in the normal state nor for a spin gap in the SC state has been
obtained [13, 14].  In contrast,
recent neutron scattering studies of optimally or slightly overdoped LSCO
have revealed a well-defined
energy gap in the incommensurate spin fluctuations below $T_c$ [14-17].
Although results of our previous
neutron scattering study indicated the existence of a spin pseudo gap at
$T_c$ for slightly overdoped LSCO
with x = 0.18 [14], no systematic examination of this system has been
performed.

In the present work, we report a comprehensive study of the magnetic
excitations in
the normal state of LSCO
with x = 0.18 and 0.20.  From a comparison of these results with
those arising from the previous
measurements on LSCO with x = 0.15 [14], we conclude that a spin pseudo
gap does exist in the LSCO system but
the visibility or the stability of the gap-like energy spectrum is
sensitively affected by the Sr or hole
concentration in these samples.

\section{Experimental Details}
\label{sec:level2}
Single crystals of $\rm La_{2-x}Sr_xCuO_4$ (x = 0.15, 0.18, 0.20) were
grown by the traveling solvent floating zone
method (TSFZ) using lamp-image furnaces [18, 19].  The as-grown single
crystals were annealed under oxygen
gas-flow at $900^{\circ}$C for 50 hrs to remove any oxygen defects.  Both x
= 0.15 and x = 0.18 crystals had been
previously used in neutron scattering measurements with some data also
being reproduced in the present
paper [14].  Onset temperatures of the SC transition measured by SQUID
magnetometers under a magnetic
field of 10 Oe are 37.5 K for the x = 0.15 sample, 36.5 K for the x = 0.18
sample, and 30.0 K for the
x = 0.20 sample (Fig. 1).  $T_c$ values for the x = 0.18 and 0.20 samples
are lower than that of the x = 0.15
sample due to overdoping.

Since $T_c$ is relatively insensitive to Sr concentration
near the optimally doped region, we
investigated the structural phase transition temperature ($T_{s}$) between the
high temperature tetragonal (HTT) and
low temperature orthorhombic (LTO) phases.  The (1,1,0) (in I4/mmm
notation) fundamental Bragg peak intensity
was monitored as a function of temperature for both x = 0.18 and 0.20
samples.  We note that the intensity
changes like an order parameter upon entering the LTO phase due to the
suppression of
extinction effects on the neutron beam caused by formation of twinned domain.
For the x = 0.15 sample, intensity of the (0,1,4) (in Bmab notation)
superlattice reflection was monitored as a function of temperature.  We
fitted the observed temperature
dependence of the peak intensity using a phenomenological function of
(1-$T/T_{s})^{2\beta}$, including a Gaussian
distribution of $T_{s}$ (the half width at half maximum of the Gaussian is
defined as $\Delta T_{s}$) to evaluate $T_{s}$
quantitatively.  The index of $\beta$ was fixed at a value of 0.35.  As a
result, $T_{s}$ and $\Delta T_{s}$ were respectively
determined to be 191 K and 10 K for the x = 0.15 sample, 111 K and 13 K for
the x = 0.18 sample, and 92 K
\begin{figure}
\includegraphics[width=\columnwidth]{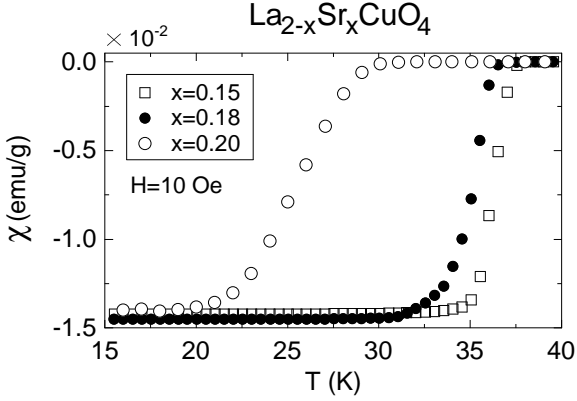}
\caption{\label{fig:squid} Shielding signals of x = 0.15 (open squares), x 
= 0.18
(closed circles) and
x = 0.20 (open circles) measured under a magnetic field of $H$ = 10 Oe.}
\end{figure}
and 22 K for the x = 0.20 sample.  Larger $\Delta T_{s}$ for the overdoped 
sample is due to larger Sr dependence of $T_{s}$.  Details of sample 
preparation and characterization have been reported
elsewhere [14, 18-20].

Inelastic neutron scattering measurements were performed using the Tohoku
University triple-axis
spectrometer TOPAN in JRR-3M of JAERI at Tokai.  The incident (final)
neutron energy was fixed at
$E_{i}$ ($E_{f}$) = 14.75 meV or 13.75 meV using the (002) reflection of a
pyrolytic graphite monochromator and
an analyzer.  The typical horizontal collimator sequence was
40'-100'-S-60'-80' or 40'-60'-S-60'-80'
where S denotes the sample position.  A pyrolytic graphite filter and a
sapphire crystal were inserted
to reduce neutron beam flux from the higher order reflection and
high-energy neutrons, respectively.
In order to increase the sample volume, two or three single crystalline
rods were
assembled and mounted in an Al container filled with He thermal exchange
gas.  A closed cycle $^{4}$He
refrigerator was used to cool samples down to 8 K with temperatures
monitored by a Si diode.
\begin{figure}
\includegraphics[width=\columnwidth]{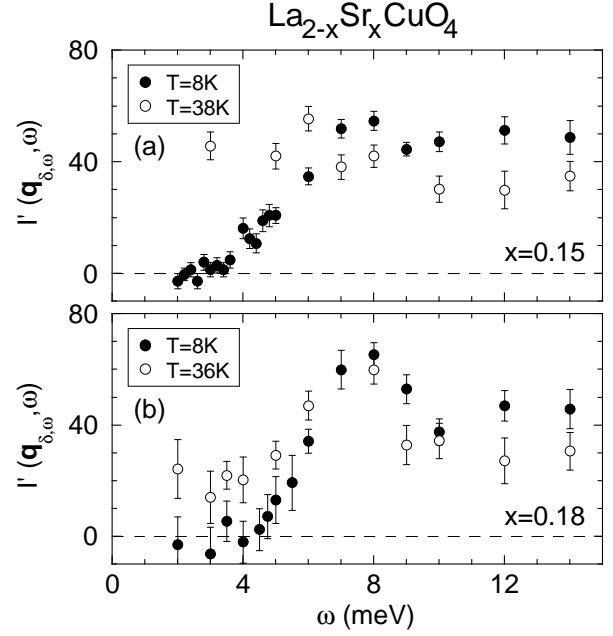}
\caption{\label{fig:peak-int} Energy dependence of the intensity at the incommensurate 
peak position for (a) x=0.15 and (b) x = 0.18 obtained at $T$ $\ll$ $T_c$ 
(closed circle) and $T$ = $T_c$ (open circle).  Data below $\omega$ = 6 meV 
for x = 0.18 were taken with $E_{i}$-fixed mode.  Data above 
$\omega$ = 6 meV for x = 0.18 and all data for x = 0.15 were taken 
with $E_{f}$-fixed mode.}
\end{figure}

\section{Analysis of Neutron Scattering Experiments}

\begin{figure}
\includegraphics[width=\columnwidth]{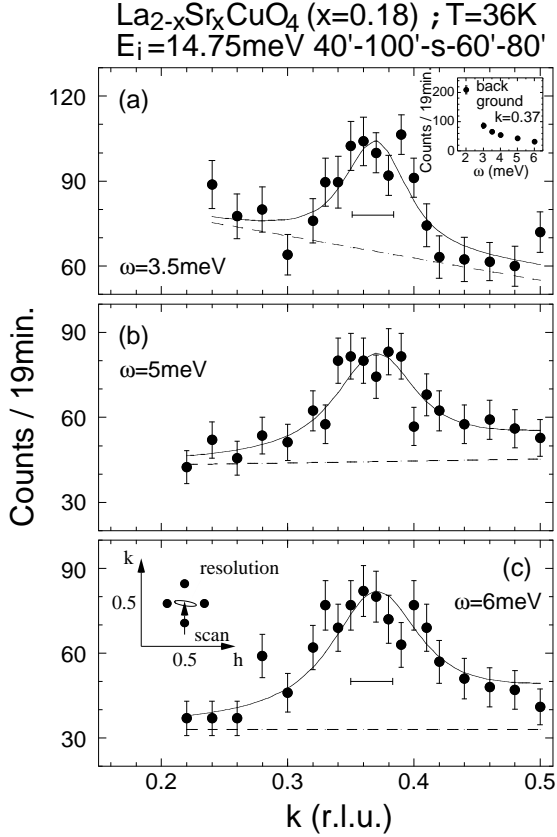}
\caption{\label{fig:q-spec18} Energy dependence of the q-spectrum of the
incommensurate magnetic
signals for x = 0.18 obtained at $T$ = $T_c$ by a constant energy-scan
around ($\pi ,\pi$).
The scan trajectory and the instrumental resolution ellipsoid are schematically shown in
the inset of (c).  
The incident neutron energy was fixed at $E_{i}$ = 14.75 meV.  Solid lines
are the results
of fits convoluted with the instrumental resolution using the background
shown by
the dashed line (see text for detail). The energy dependence of back ground 
is depicted in the inset of (a). 
Horizontal bars depict the instrumental
q-resolution. The intensity at around ($\pi ,\pi$), $k$ = 0.5 in the figure,
reflects the
effect of incommensurate peaks located outside of the scan trajectory (see
inset of (c)).}
\end{figure}
The energy dependence of incommensurate peak intensity is depicted (Fig. 2) after 
making following correction on raw data arised from instrument. 
Background-subtracted inelastic scattering intensities, $I$, taken with 
$E_{i}$-fixed mode are corrected into $I'$ using 
the following equation,

\begin{equation}
I' = I \cdot \frac{tan\theta_{A}}{k_{f}^3}
\end{equation}

\noindent where $k_{f}$ and $\theta_{A}$ denote 
wavenumber of the scattered neutrons and the analyzer angle, respectively [21].  
On the other hand, for experiments with 
$E_{f}$-fixed mode, the counting time was corrected.  This was necessary 
as the count rate of a fission monitor for the incident beam flux 
depends on the incident neutron energy due to the energy dependence of 
intensity of the higher order reflected beam.

For quantitative analysis, we fitted the observed magnetic intensity, which
is proportional to the
dynamical structure factor, $S({\bf q},\omega)$, and dynamical magnetic
susceptibility, $\chi^{\prime\prime}({\bf q},\omega)$, using the
following equations convoluted with an instrumental resolution function,

\begin{equation}
S({\bf q},\omega)=\frac{1}{1-\exp(-\frac{\omega}{k_{\rm B}T})} \cdot
\chi^{\prime
\prime}({\bf q}, \omega)
\end{equation}

\begin{equation}
\chi^{\prime\prime}({\bf q},\omega) = A_\omega \sum_{\delta=1,4} \left\{
\frac{\kappa_\omega}{|{\bf q}-{\bf q}_{\delta,\omega}|^2+\kappa^2_\omega} \right\}
\end{equation}

\noindent where $k_{\rm B}$, ${\bf q}_{\delta,\omega}, \kappa_\omega$ and 
$A_\omega$ denote 
the Boltzman constant, the four-fold positions of incommensurate
peaks around ($\pi ,\pi$), the q-width and a scaling factor, respectively.
The absolute values of q-integrated dynamical magnetic susceptibilities,
$\chi^{\prime \prime} (\omega)$, are determined using phonon intensities as 
described previously [14].

\section{Results}
\label{sec:level3}
\begin{figure}
\includegraphics[width=\columnwidth]{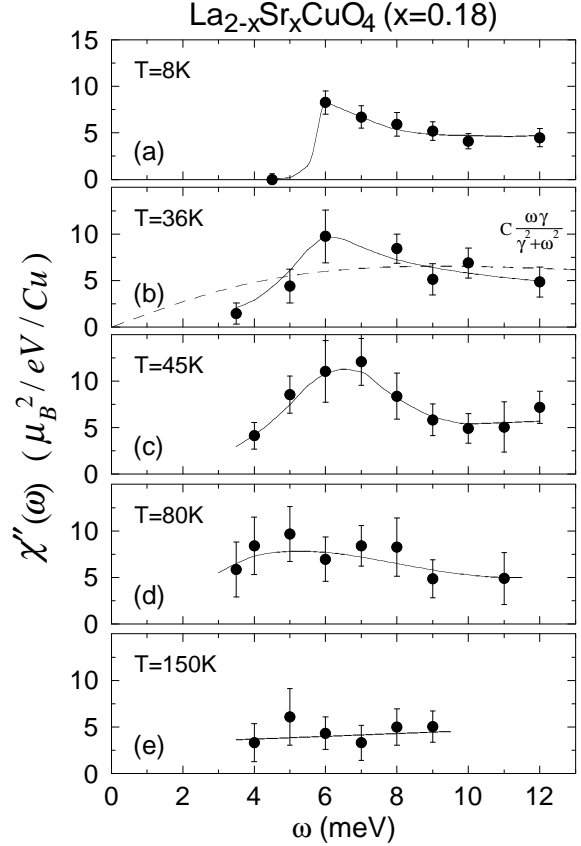}
\caption{\label{fig:kai18} Energy spectrum of $\chi^{\prime \prime} 
(\omega)$ for the x = 0.18
sample in the temperature range 8 K $\leq$ $T$ $\leq$ 150 K.  
Data below $\omega$ = 6 meV at $T$ = 8 and 36 K were taken 
with $E_{i}$-fixed mode and the others with $E_{f}$-fixed mode.  
At $T$ = 36 and 45 K, a gap-like
energy spectrum remains.  Solid lines provide guides to the eyes.  
A dashed line is a fit using Eq. (4).}
\end{figure}
Figure 2 shows energy dependence of incommensurate peak intensity for 
x = 0.15 and 0.18.  At $T$ $\ll$ $T_{c}$, both samples have a clear gap 
spectrum as reported in ref. [14, 15].  
Upon heating to $T$ = $T_{c}$, on the other hand, the gap structure 
is disappeared for x = 0.15.  
Whereas, for x = 0.18, although the intensity does not drop into 0, reduction 
of peak intensity at low energy region still occurs, which suggests that a
spin pseudo gap is open.  Note that the intensity below $\omega$ = 6 meV 
increases with increasing temperature while it decreases above $\omega$ = 6 meV, 
which can be owing to sum rule.  

Typical raw q-spectra of magnetic peaks for the x = 0.18 sample taken by the
$E_{i}$-fixed mode at $T$ $\sim$ $T_c$
are shown in Fig. 3.  Trajectory of the scan is illustrated in the inset of Fig. 3(c).
Solid lines depict the
results of fits using the scattering function described in equation (2) and
(3) convoluted with
the resolution function.  Long dashed lines depict background.  As shown,
the solid lines reproduce
the observed q-spectra quite well.  Larger magnetic intensity near the
($\pi ,\pi$) position at $\omega$ = 6 meV than the left
side tail dominantly arises from incommensurate peaks
outside of the scan trajectory collected by the finite instrumental
resolution (see inset of Fig. 3(c)).  
Energy dependence of back ground at incommensurate peak position ($k$ = 0.37 in Fig. 3) 
is depicted in inset of Fig. 3(a).  
Back ground at $\omega$ = 3.5, 5 and 6 meV is estimated from least square fits 
as shown in Fig. 3.  Whereas at $\omega$ = 2, 3 and 4 meV, it is estimated by 
taking an average at both side of peak tail.  
The smooth energy dependence suggests 
that the present back ground is reasonable.  
Furthermore we note that if 10 counts larger back ground is used 
for the data at for example $\omega$ = 6 meV 
(Fig. 3(c)), no more reasonable fitting line can be obtained.

Figure 4 shows energy dependence of $\chi^{\prime \prime} (\omega)$ for the
x = 0.18 sample.
As presented in Fig. 4(a) and
in our previous paper [14], the energy spectrum at $T$ = 8 K exhibits a
clear cut-off near $\omega$ = 6 meV ($E_{gap}$) with a broad bump, 
while $\chi^{\prime \prime} (\omega)$ completely vanishes into
the background below $\omega$ $\sim$ 4.5 meV.   Upon heating the
sample to $T_c$, magnetic intensities are found to
appear below $\omega$ = 4.5 meV
(Figs. 2(b) and 3(a)).  Nevertheless, steep decrease in 
$\chi^{\prime \prime} (\omega)$ below $\omega$ = 6 meV 
remains, suggesting a presence of spin pseudo gap.  
The bump around $\omega$ = 6 meV is also survived.  
Dashed line in Fig 4(b) depicts a fit assuming a gapless state using Eq. (4), 
which details are described in section 5.  
Upon heating the sample to $T$ = 80 K, the gap-like structure becomes
broad, while at $T$ =
150 K, $\chi^{\prime \prime} (\omega)$ is nearly independent of
temperature, consistent with a collapse of the spin pseudo gap.

Fig. 5 shows the energy dependence of line width.  
At $T$ = 8 K (Fig. 5(a)), the line width 
\begin{figure}
\includegraphics[width=\columnwidth]{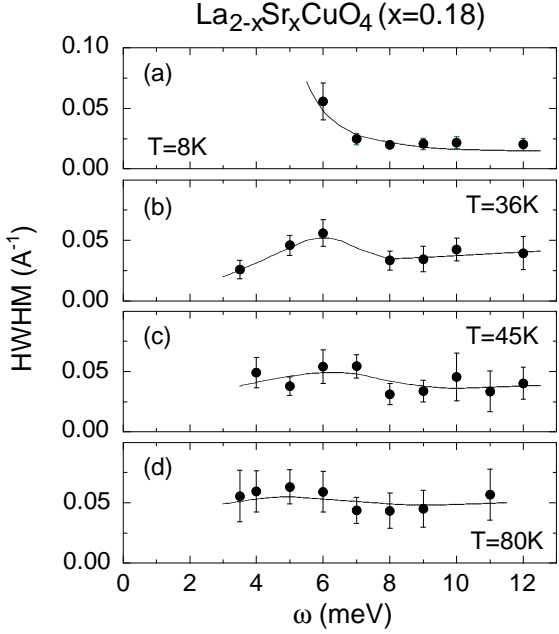}
\caption{\label{fig:line} Energy dependence of the resolution deconvoluted
line-width of an
incommensurate peak for the x = 0.18 sample in the temperature range  8 K
$\leq$ $T$ $\leq$ 80 K.
At $T$ = 8 K, due to the opening of energy gap, the line-width is not defined
below the gap energy.
Solid lines provide guides to the eyes.}
\end{figure}
\begin{figure}
\includegraphics[width=\columnwidth]{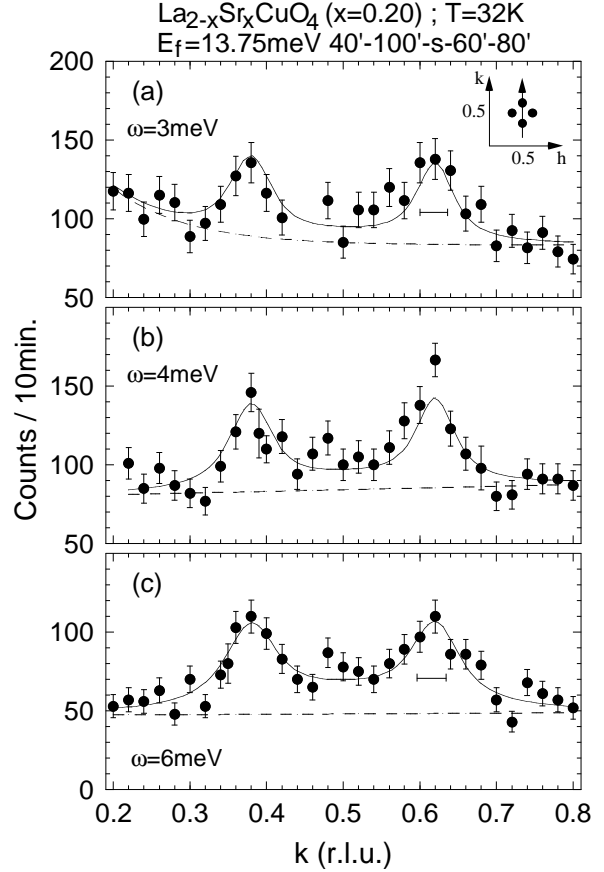}
\caption{\label{fig:q-spec20} Energy dependence of the q-spectrum of 
incommensurate
magnetic
signals for x = 0.20 at $T$ = $T_c$.  The final neutron energy was fixed at
$E_{f}$ = 13.75 meV.  Solid lines depict the results of fits convoluted
with the instrumental resolution using the background shown by the dashed 
lines.}
\end{figure}
\begin{figure}
\includegraphics[width=\columnwidth]{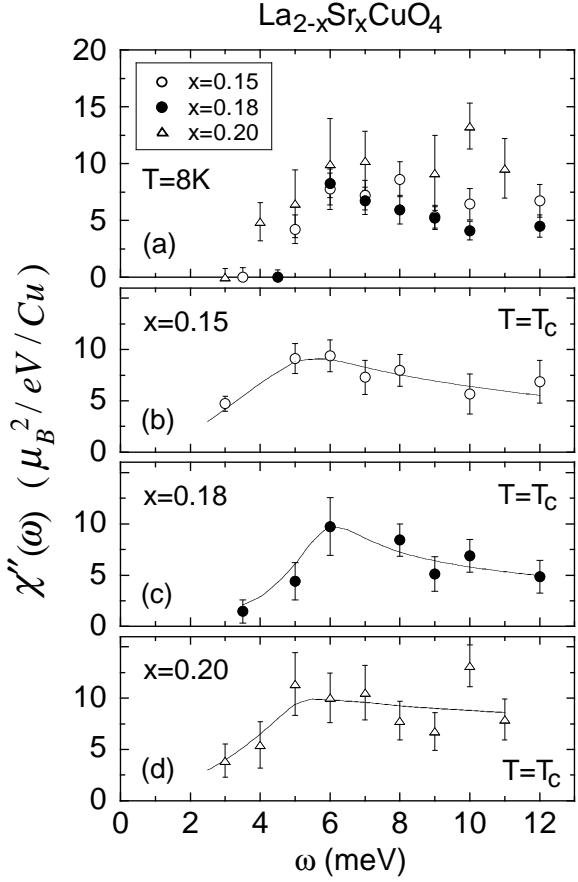}
\caption{\label{fig:kai-sum} Energy dependence of $\chi^{\prime \prime} 
(\omega)$ for
the x = 0.15, 0.18 and 0.20 samples at (a) $T$ = 8 K and (b - d) $T$ =
$T_c$.  At $T$ = 8 K three samples shows clear energy gap with the
gap-energy around 6 meV. At $T$ = $T_c$, only the x = 0.18 sample shows a
gap-like energy spectrum.  Solid lines provide guides to the
eyes.}
\end{figure}
increases with decreasing the energy down to $\omega$ = 6 meV.  
Below the energy gap, the line width cannot be defined due to
the absence of magnetic intensity.  
At $T$ = 36 K, the enhancement around $\omega$ = 6 meV still remains 
as a small bump (Fig. 5(b)).  On the other hand, 
upon heating to $T$ = 80 K, the width becomes nearly independent of energy.
It seems that the bump near the energy gap disappears with the 
closing of the spin pseudo gap.  Possibly, the bump is an effect of 
the spin pseudo gap as discussed in the previous paper [14].

Figure 6 shows the raw q-spectrum of magnetic peaks for the x = 0.20 sample at
$T$ $\sim$ $T_c$ taken at $\omega$ = 3, 4
and 6 meV under the $E_{f}$-fixed mode. Well-defined incommensurate peaks
with the peak-width similar
to that of the x = 0.18 are observed.  We find that the peak intensity has
a weaker
energy dependence compared to the x = 0.18 sample. Note that a clear
energy gap exists for $T$ $\ll$ $T_c$ (see Fig. 7(a)) with the gap energy of
5 $\sim$ 6 meV, which is slightly smaller than that
of the x = 0.18 sample.  At $T$ $\sim$ $T_c$, as shown in the Fig. 7(d), however, no clear 
gap-like structure is observed, whereas a weak bump is seen at 
$\omega$ $\sim$ $E_{gap}$.

Fig. 7 summarizes the energy dependences of $\chi^{\prime \prime} (\omega)$ for
the x = 0.15 (data from ref. [14]),
0.18 and 0.20 samples.  At $T$ = 8 K, all three samples show a well-defined
energy gap with a broad
maximum near $\omega$ = 6 meV.  Upon heating to $T$ = $T_c$, a gap-like
structure of $\chi^{\prime \prime} (\omega)$ remains only for the x = 0.18
sample.  For the x =
0.15 and 0.20 samples, $\chi^{\prime \prime} (\omega)$ decreases linearly
with decreasing energy below $\omega$ $\sim$ 5 meV, approaching zero only
as $\omega$ $\rightarrow$ 0.  The broad peak in $\chi^{\prime \prime} 
(\omega)$,
on the other hand, is still observed in all three samples.

\section{Discussion}
\label{sec:level4}
A signature of spin pseudo gap in the energy
spectrum of incommensurate spin
fluctuations was first obtained for the 2-1-4 type hole-doped cuprates.
The present neutron scattering experiment
shows that a gap-like structure at $T$ $\sim$ $T_c$ is observable only in a 
narrow Sr concentration range near
x = 0.18.  For the x = 0.15 and 0.20 samples, although a clear gapped
spectrum was observed at $T$ $\ll$ $T_c$,
the spin pseudo gap was poorly defined at $T$ $\sim$ $T_c$.
\begin{figure}
\includegraphics[width=\columnwidth]{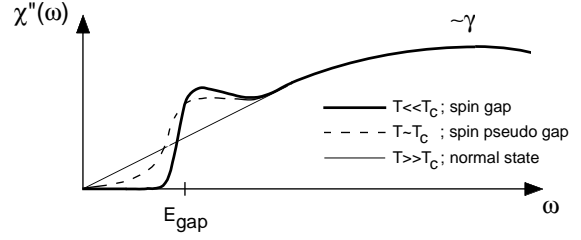}
\caption{\label{fig:kaidraw} A  conceptual drawing of the q-integrated 
dynamical
magnetic susceptibility $\chi^{\prime \prime} (\omega)$ in a high-$T_c$ cuprate
system with a spin-gap, a spin pseudo gap and without a gap.  In the normal state well 
above $T_c$,
$\chi^{\prime \prime} (\omega)$ exhibits a broad peak near $\gamma$, a 
characteristic
energy-scale of spin fluctuations (see text and Eq. (4)), and depends 
linearly on $\omega$
near $\omega$ = 0.  In the superconducting state well below $T_c$, 
$\chi^{\prime \prime} (\omega)$
shows a fully opened gap accompanied by a broad bump near the gap energy
(spin-gap shown by the thick solid line).  In the intermediate temperature 
region,
we can expect a spin pseudo gap state as shown by the broken line in the figure
where downward deviation from the linear dependence and the broad bump are observed but the gap is 
not fully opened.}
\end{figure}

We first discuss the
broad peak in $\chi^{\prime \prime} (\omega)$ with $\omega$ $\sim$ 6 meV 
in view of the spin pseudo gap.  
In general, 
for the correlated spin systems without magnetic long range order, 
$\chi^{\prime \prime} (\omega)$
exhibits a broad peak even in the normal state.  For example, 
$\chi^{\prime \prime} (\omega)$ can be described in a Lorentzian form

\begin{equation}
\chi^{\prime \prime}(\omega) = C \frac{\omega \gamma}{\gamma^2+\omega^2} \ \ ,
\end{equation}

\noindent where the peak energy nearly 
corresponds to $\gamma$, a characteristic energy-scale of spin fluctuations 
of the system.  
Thus, the 
broad peak in $\chi^{\prime \prime} (\omega)$ does not simply correspond to
a spin gap or spin pseudo gap.
For the optimally doped LSCO, however, $\chi^{\prime \prime} (\omega)$ has
already been studied over a wide energy range by pulse
neutron scattering and a broad peak was observed
at $\omega$ = 22 $\sim$ 40 meV [22-24].  
Then, combining both results from the present low energy experiments
and pulse neutron scattering at high energy, there should exist 
two peaks in $\chi^{\prime \prime} (\omega)$.  
A poorly fit using Eq. (4) to $\chi^{\prime \prime} (\omega)$ of x = 0.18 at $T$ = 36 K 
where $\gamma$ = 9 meV is obtained (Fig. 4(b)) suggests that the 
$\chi^{\prime \prime} (\omega)$ at low energy region can not be explained by the
simple Lorentizan form.  
Possibly, the peak at lower energy is the result of forming an energy
gap below $T_c$ and even a spin pseudo gap at $T$ $\sim$ $T_c$.  
Fig. 8 depicts a conceptual drawing of $\chi^{\prime \prime} (\omega)$.  
In normal state, $\chi^{\prime \prime} (\omega)$ depends linearly on $\omega$ 
at low energies near $\omega$ = 0.  If spin gap or spin pseudo gap opens, 
the $\chi^{\prime \prime} (\omega)$ near 0 deviates downward from the linear 
dependence and a bump appears near the gap - energy.

According to the fact that the broad peak in $\chi^{\prime \prime} (\omega)$ 
with $\omega$ $\sim$ 6 meV is also observed in x = 0.15 and 0.20, we speculate 
that the spin pseudo gap remains at $T$ $\sim$ $T_c$ in these samples although 
steep decrease of $\chi^{\prime \prime} (\omega)$ is less defined.  
We note that the visibility or the 
stability of the pseudo gap highly depends on the gap-edge structure of the 
energy spectrum at $T$ $\ll$ $T_c$, which is broader at x = 0.15 and 0.20 than at x 
= 0.18.  
If gap-edge is broader at base temperature, gap-like structure or spin pseudo
gap is more easily smeared out by heating. Then the spin pseudo gap is difficult
to detect even if it exists.
One possible reason for the gap-edge broadening in the ground state is the
spatial distribution of the size of energy gap due to the inhomogeneous
carrier distribution and/or chemical potential-randomness. The effect of
the randomness becomes dominant below x $\sim$ 0.15 due to the substantial
decrease in the effective carrier concentration [25].  
For the x = 0.20 sample, on the other hand, the spin pseudo gap is possibly 
degradaded by overdoping as reported by many other indirect experiments.  
At present, however, the direct relation between the stability of spin 
pseudo gap and the gap-edge broadening is not clear.

Many studies on pseudo gap have been predicting the existence of two
characteristic crossover temperatures, $T_{\rm o}$
and $T^{\rm *}$ ($T_{\rm o}$ $>$ $T^{\rm *}$).
For example, at both temperatures of $T_{\rm o}$ and $T^{\rm *}$,
suppression of magnetic susceptibility and downward deviation of inplane 
resistivities
from linear temperature dependence are observed [26-28].
Furthermore, at $T^{\rm *}$, the electronic specific-heat coefficient is 
slightly suppressed [29].
The origin of such $T_{\rm o}$
is interpreted by the onset of
antiferromagnetic correlation.  Below $T^{\rm *}$, on the other hand, the
pseudo gap is expected to open.  According to NMR measurements, 
the value of $T^{\rm *}$ for x = 0.20 is about $T$ = 200 K [7].  
Taking into account the difficulty of determining $T^{\rm *}$ correctly, 
the thermal scale of the spin pseudo gap observed in the present study
appears to be consistent with the results of NMR.

The relation between the pseudo gap and superconducting gap is also of
great interest.  According to
angle-resolved photoemission measurements in LSCO, large and small charge
pseudo gaps exist with magnitudes
of about 100 meV and 30 meV, respectively [4, 5].  The present results,
however, suggest that a single spin
gap structure in the $\chi^{\prime \prime} (\omega)$ spectrum at base
temperatures gradually transforms into the spin pseudo gap upon heating.

\section{Conclusion}
\label{sec:level5}
We observed a signature of a spin pseudo gap in the energy spectrum of
$\chi^{\prime \prime} (\omega)$
for the slightly overdoped x = 0.18 samples of LSCO by neutron scattering
experiments.  Upon
heating, the spin pseudo gap gradually collapses between $T$ = 80 K and 150 K.  
For the x = 0.15 and x = 0.20 samples, on the other hand, the gap-like
structure is poorly
defined at $T$ = $T_c$.  However, the broad bump in $\chi^{\prime \prime}
(\omega)$ at $T$ $\sim$ $T_c$
suggests the remaining of the spin pseudo gap.

\begin{acknowledgments}
The authors would like to appreciate M. Matsuda , J. M. Tranquada and Y. S. Lee
for valuable discussions.  The work was supported by a
Grant-In-Aid for Scientific Research on Priority Areas, Novel Quantum
Phenomena in Transition Metal
Oxides and for Scientific Research (A) from the Ministry of Education,
Culture, Sports, Science and
Technology of Japan and by a Grant from the Japan Science and Technology
Corporation, the Core Research
for Evolutional Science and Technology Project and by a Grant from the
Ministry of Economy, Trade and Industry of Japan.
\end{acknowledgments}

\end{document}